# The Fulde-Ferrell-Larkin-Ovchinnikov State in the Organic Superconductor κ-(BEDT-TTF)$_2$Cu(NCS)$_2$ as Observed in Magnetic Torque Experiments

B. Bergk[a], A. Demuer[b], I. Sheikin[b], Y. Wang[c], J. Wosnitza[a], Y. Nakazawa[d], and R. Lortz[c]

[a]*Hochfeld-Magnetlabor Dresden, Forschungszentrum Dresden-Rossendorf, 01314 Dresden, Germany*

[b]*Grenoble High Magnetic Field Laboratory, CNRS, 25 avenue des Martyrs, BP 166, 38042 Grenoble*

[c]*Department of Physics, The Hong Kong University of Science & Technology, Clear Water Bay, Kowloon, Hong Kong*

[d]*Department of Chemistry, Osaka University, 1-1, Machikaneyama, Toyonaka, Osaka, Japan*

**Abstract**

We present magnetic-torque experiments on the organic superconductor κ-(BEDT-TTF)$_2$Cu(NCS)$_2$ for magnetic fields applied parallel to the 2D superconducting layers. The experiments show a crossover from a second-order to a first-order transition when the upper critical field reaches 21 T. Beyond this field, which we interpret as the Pauli limit for superconductivity, the upper critical field line shows a pronounced upturn and a phase transition line separates the superconducting state into a low- and a high-field phase. We interpret the data in the framework of a Fulde-Ferrell-Larkin-Ovchinnikov state.

## 1. Introduction

The upper critical field of type-II spin-singlet superconductors is usually determined by the orbital limit for superconductivity, where the current density around the vortex cores reaches a pair breaking value and the normal state is restored. In layered superconductors the open nature of the Fermi surface raises this limit to very high in-plane magnetic fields. The orbital limit may then exceed the Pauli limit ($H_P$), above which the Zeeman-split Fermi surfaces no longer allow Cooper pairing with zero center-of-mass momentum. The theory of P. Fulde and R.A. Ferrell as well as A.I. Larkin and Y.N. Ovchinnikov [1] predicts that type-II superconductors have the possibility to increase their upper critical fields beyond $H_P$ by forming the Fulde-Ferrell-Larkin-Ovchinnikov (FFLO) state. This state is characterized by a finite center-of-mass momentum of the Cooper pairs, which results in an oscillating part of the order parameter in real space, with wavelength of the order of the coherence length ξ. One candidate for the appearance of the FFLO state is the heavy-fermion compound CeCoIn$_5$ [2,3]. The exact nature of the high-field state in CeCoIn$_5$ remains controversially debated [4]: A coupling between superconducting and magnetic order has been revealed by neutron-scattering experiments [5]. Layered organic compounds have been discussed as further promising superconductors to develop the FFLO state [6,7,8]. Many of them are non-magnetic which means that any involvement of magnetism is unlikely. Indeed, some signs for an FFLO state in 2D [9,10,11] and 1D [12,13] organic superconductors have been reported. We reported an earlier evidence from specific-heat experiments for the FFLO state in κ-(BEDT-TTF)$_2$Cu(NCS)$_2$ (where BEDT-TTF is bisethylenedithio-tetrathiafulvalene) [14]. In the present article, we present magnetic-torque data in order to confirm this result by a second thermodynamic quantity.

## 2. Experiment

Single crystals of this layered organic charge transfer salt were grown by the standard electrochemical oxidation method as described elsewhere [15]. Measurements of the magnetic torque were performed using a capacitance cantilever technique in a 28 as well as in a 32 T resistive magnet of the Grenoble High Magnetic Field Facility. The experi-mental probe was equipped with a rotator, which allowed us to align the superconducting layers parallel to the magnetic field direction with a precision of 0.001°. A measurement of de-Haas-van-Alphen oscillations of the samples in the normal state at a field orientation close to 45° indicated that the sample is in the clean limit. Fig. 1a shows magnetic-torque data taken at various fixed temperatures.

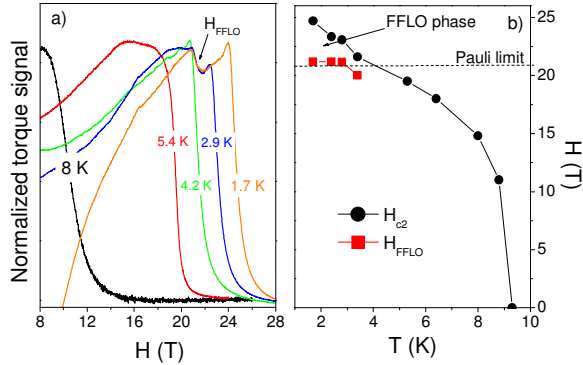

Figure 1. a) Magnetic torque of κ-(BEDT-TTF)$_2$Cu(NCS)$_2$ for in-plane magnetic fields at 8, 5.4, 4.2, 2.9, and 1.7 K (from left to right). b) Magnetic phase diagram constructed from the data.

The first observation is that the continuous kink-like second-order transition at $H_{c2}$, as observed at 8 K, changes into a step-like first-order transition at temperatures below ~5.5 K. This is a clear sign that $H_P$ is reached at ~ 21 T in accordance with earlier observations [14]. However, superconductivity survives beyond the Pauli limit and a small downward step-like feature indicates a thermodynamic transition into a novel superconducting high-field state. This transition line within the superconducting state always remains close to $H_P$. We choose the inflection point in the torque signal as criteria to define the transitions and plot them in a $H$-$T$ phase diagram (Fig. 1b). The large initial slope of the $H_{c2}$ line at zero field starts to saturate at 21 T, exactly where the crossover to first-order nature occurs. Beyond 21 T, a pronounced upturn indicates an abrupt change in the condensate below $H_{c2}$. It is rather obvious that this change is related to the phase transition within the superconducting state, as it approaches the $H_{c2}$ line at the onset of the upturn.

## 3. Discussion

For singlet pairing, superconductivity beyond $H_P$ is only possible in terms of a finite center-of-mass momentum of the Cooper pairs. Remarkably, the second transition within the superconducting state remains close to $H_P$, where the transition from a state with zero momentum Cooper pairs to pairs with finite momentum must occur. The drop of the magnetic torque at this field indicates that a part of the condensate is lost above. This is expected upon the development of a spatial modulation of the order parameter. Some hysteresis has been observed between the curves measured upon ascending and descending field (data not shown) indicating metastability, presumably related to the first-order nature of the transition.

The main features of the phase diagram nicely confirm our previous specific-heat results. However, there is a significant difference in the phase diagram obtained with the two methods: The specific-heat data showed the FFLO transition line closely following the $H_{c2}$ line, whereas in the torque data the line remains at nearly constant field close to $H_P$. In order to explain this difference, one needs to point out the different conditions in the two experiments: Specific heat was measured in constant fields, whereas torque was measured at fixed temperatures during field sweeps. The observed metastability may then lead to different locations of the FFLO transition lines in the two experiments. Furthermore, in the specific-heat experiments the alignment of the sample was not possible better than ~1°. The torque sensor mounted on a rotator allowed us to align the sample much more precisely. A better parallel alignment may favor a larger FFLO phase.

## 4. Conclusion

From our magnetic torque data, we obtain a high magnetic field phase diagram of κ-(BEDT-TTF)$_2$Cu(NCS)$_2$ for in-plane field orientations with a clear signature of an unconventional superconducting high-field phase. Magnetic order can be excluded for κ-(BEDT-TTF)$_2$Cu(NCS)$_2$, therefore this phase most probably represents the true FFLO state as predicted by P. Fulde, R.A. Ferrell, A.I. Larkin and Y.N. Ovchinnikov [1].